\documentclass[11pt,preprint]{aastex}
\usepackage{epstopdf}
\usepackage{chngpage}
\usepackage{graphicx}
\usepackage{natbib}

\shorttitle{RELATIVISTIC BLAST WAVE}
\shortauthors{Geng et al.}

\begin{document}

\title{REVISITING THE EMISSION FROM RELATIVISTIC BLAST WAVES IN A DENSITY-JUMP MEDIUM}

\author{J. J. Geng\altaffilmark{1, 2}, X. F. Wu\altaffilmark{3, 4, 5}, Liang Li\altaffilmark{6, 7}, Y. F. Huang\altaffilmark{1, 2}, and Z. G. Dai\altaffilmark{1, 2}}

\altaffiltext{1}{School of Astronomy and Space Science, Nanjing University, Nanjing 210093, China; hyf@nju.edu.cn}
\altaffiltext{2}{Key Laboratory of Modern Astronomy and Astrophysics (Nanjing University), Ministry of Education, China}
\altaffiltext{3}{Purple Mountain Observatory, Chinese Academy of Sciences, Nanjing 210008, China; xfwu@pmo.ac.cn}
\altaffiltext{4}{Chinese Center for Antarctic Astronomy, Chinese Academy of Sciences, Nanjing 210008, China}
\altaffiltext{5}{Joint Center for Particle Nuclear Physics and Cosmology of Purple Mountain Observatory-Nanjing University, Chinese Academy of Sciences, Nanjing 210008, China}
\altaffiltext{6}{Department of Physics, Stockholm University, AlbaNova, SE-106 91 Stockholm, Sweden}
\altaffiltext{7}{Erasmus Mundus Joint Doctorate in Relativistic Astrophysics}

\begin{abstract}
Re-brightening bumps are frequently observed in gamma-ray burst (GRB) afterglows. Many scenarios have been proposed to interpret the origin of these bumps, of which a blast wave encountering a density-jump in the circumburst environment has been questioned by recent works. We develop a set of differential equations to calculate the relativistic outflow encountering the density-jump by extending the work of Huang et al. (1999). This approach is a semi-analytic method and is very convenient. Our results show that late high-amplitude bumps can not be produced under common conditions, only short plateau may emerge even when the encounter occurs at early time ($< 10^4$ s). In general, our results disfavor the density-jump origin for those observed bumps, which is consistent with the conclusion drawn from full hydrodynamics studies. The bumps thus should be due to other scenarios.
\end{abstract}


\section{INTRODUCTION}
\label{sect:intro}
It is believed that gamma-ray bursts (GRBs) are produced by either the collapse of massive stars (Woosley 1993; MacFadyen \& Woosley 1999) or the merger of compact binaries (Eichler et al. 1989), which can launch a collimated relativistic blast wave into circumburst medium. As the relativistic forward shock propagates into the surrounding medium, the circumburst electrons would be swept up and accelerated. These shocked electrons power the afterglow emission via synchrotron radiation, which can be observed for weeks at X-ray and optical frequencies. In the pre-{\it Swift} era, many afterglow light curves showed a smooth power-law decay. On the basis of the self-similar Blandford-McKee solution (BM: Blandford \& McKee 1976) to the dynamics of relativistic blast wave, a standard model of forward shock afterglow was developed (e.g., M{\'e}sz{\'a}ros \& Rees 1997; Panaitescu et al. 1998; Sari et al. 1998; see Zhang \& M{\'e}sz{\'a}ros 2004 and Gao et al. 2013 for recent reviews) and is generally consistent with the observations. However, some optical afterglows have shown significant temporal variabilities, with strong deviation from the typical power-law behavior (Frontera et al. 2000). This issue has long been debated among researchers.

Recent observations by {\it Swift}/XRT (Gehrels et al. 2004) reveal that early X-ray flares are common in afterglows. These early flares are probably due to the early activities of the central source (Burrows et al. 2005; Nousek et al. 2006). An overview of the optical afterglow samples (Liang et al. 2013) shows that many afterglows have optical bumps at a late time (usually $10^4$ -- $10^5$ s after the trigger), while there are no corresponding significant signal at X-ray band. To explain such temporal variabilities in GRB afterglow light curves, several scenarios have been proposed. One proposal involves the ``re-freshed shocks'', including the late energy injection model (Zhang \& M{\'e}sz{\'a}ros 2002; Kong \& Huang 2010; Geng et al. 2013) or the later internal shock model (Rees \& M{\'e}sz{\'a}ros 1994; Fan \& Wei 2005; Yu \& Dai 2009; Wang \& Cheng 2012). Notable bumps with steep rising slope in optical afterglow of some GRBs, like GRB 081029 (Nardini et al. 2011a) and GRB 100621A (Greiner et al. 2013) may be caused by this mechanism. The two-component jet model can also cause a late bump (Huang et al. 2006), and it has been well applied to some cases (e.g., GRB 030329, Berger et al. 2003). Alternatively, a possible origin for variabilities in the afterglows is blast wave encountering a jump in density.

Density fluctuations near the GRB progenitor are expected because of strong winds and violent mass eruptions (Ramirez-Ruiz et al. 2001) prior to the collapse of its core. Thus, many authors studied the interaction of a blast wave with density structures to check whether the density-jump (or decrease) can explain the complex variations in the afterglows. Dai \& Lu (2002) have calculated the emission when a relativistic blast wave interacts with a density-jump medium. In their analytical solution, a notable bump appears in the afterglow if a resultant reverse shock is relativistic. Dai \& Wu (2003) applied this analytical solution to GRB 030226 to explain a rebrightening bump in this GRB afterglow. Other detailed numerical calculations (Lazzati et al. 2002; Mesler et al. 2012) show similar results and are applicable to some specific afterglows. Uhm \& Zhang (2014) also pointed that some observed features (such as re-brightenings, dips, or slow wiggles) can be explained with the density fluctuations. However, the results from van Eerten et al. (2009, 2010) and Nakar \& Granot (2008) are different: they suggest that the density-jump does not produce sharp flares in the light curves. Most recently, Gat et al. (2013) gave an analytical solution and also operated a numerical simulation to this problem, confirming the study of van Eerten et al. (2009). Thus the contradiction between the analytical result and full hydrodynamics simulation motivated us to revisit this issue with a further semi-analytic numerical approach.

The dynamics of a forward shock surfing in a smooth density profile environment can be well described by the differential equations proposed by Huang et al. (1999, 2000a, 2000b). In this paper, we extend the derivation of Huang et al. (1999) to include the role of the reverse shock emerging during the interaction. Our results can help to explore whether the onset of the emission from the reverse shock can lead to the late bumps in the optical light curves. Our paper is organized as follows. In Section 2, we analyze the hydrodynamics of the blast wave expanding in the density-jump medium. In Section 3, we present the formula for the synchrotron emissivity. We present our numerical results and the comparison with previous work in Section 4. Our conclusions are summarized in Section 5. Details of our derivations for the dynamic equations are described in Appendix A and Appendix B.

\section{HYDRODYNAMICS}
\label{sect:dyna}
After the prompt emission of GRBs, the blast wave will sweep up the ambient medium. We assume that the medium has a simple number density profile as a step function of radius $R$: $n = n_0$ for $R \leq R_0$ and $n = n_1$ for $R \geq R_0$ ($n_1 > n_0$), where $R_0$ is the transition radius. If the density contrast is higher than some threshold, the interaction of the forward shock with the high-density medium will lead to a reverse shock propagating into the hot shell (Sari \& Piran 1995; Kobayashi et al. 1999; Kobayashi \& Sari 2000). Thus the overall evolution of the blast wave should be divided into three episodes: the stage before encountering the density-jump, the period when the reverse shock is crossing the shell, and that after the reverse shock crosses the shell. We will carefully analyze these episodes respectively below.

\subsection{Before Encountering the Density Jump}
Let's consider a forward shock propagating into the cold interstellar medium (ISM). The evolution of the forward shock is calculated by considering energy conservation. We assume the initial mass of the outflow is $M_{\rm ej}$. The shock front separates the system into two regions: (1) the unshocked ISM, (2) the shocked ISM. We treat these regions as simple homogenous shells (see Piran 1999 for an exhaustive treatment of this approach). If the outflow propagates with Lorentz factor $\gamma_2$ at time $t$, the thermodynamical quantities of the gas in the rest frame of region 2 are given by: $U_2'=\psi_2 m_2 c^2=(\gamma_2-1)m_2 c^2$ (internal energy) and $P_2^{\prime}=(\hat{\gamma}-1) U_2^{\prime} =\frac{\psi_2+2}{3(1+\psi_2)}\psi_2m_2c^2$ (pressure), where $m_2$ is the total mass swept up by the shell and $\hat{\gamma}$ is the adiabatic index. Here and below, we use prime ($\prime$) on variables to denote quantities in the shock comoving frame and characters without a prime to denote quantities in the observer frame. If a fraction $\epsilon_2$ (radiation efficiency) of the thermal energy is radiated, then the ``total'' energy (not including the rest mass energy) of the gas is given by $E_2 \simeq (\gamma_2-1)(m_2+M_{\rm ej})c^2+(1-\epsilon_2)\gamma_2(U_2^{\prime}+P_2^{\prime})$. In fact, a more exact expression for the total energy is $E_2 =\left\{\gamma_2+(1-\epsilon_2)\left[\hat{\gamma}\gamma_2^2-\gamma_2-(\hat{\gamma}-1)(1+\gamma_2\beta_2^2)\right]\right\}m_2 c^2$ (Pe'er 2012), where $\beta_2$ is the velocity of region 2 and $c$ is the light velocity. We use the approximation just for the simplification of the equations derived below, and this is especially helpful for the treatment of the episode in the next subsection. Since the shock-accelerated
electrons carry only a fraction $\epsilon_e$ of the internal energy, the radiation efficiency of the total ejecta can be given by $\epsilon_2=\epsilon_e {t_{\rm syn}^{\prime}}^{-1}/({t_{\rm syn}^{\prime}}^{-1}+{t_{\rm ex}^{\prime}}^{-1})$ (Dai et al. 1999), where $\epsilon_e$ is the equipartition parameter for electrons, $t_{\rm syn}^{\prime}$ is the synchrotron cooling time and $t_{\rm ex}^{\prime}=R/(\gamma_2c)$ is the comoving frame expansion time.

Following the procedure of Huang et al. (1999), we can derive the differential equation for the evolution of the Lorentz factor $\gamma_2$. When the ISM of mass $dm_2$ is swept up by the shell, a fraction $\epsilon_2$ of the gained thermal energy is assumed to be radiated, thus the lost energy is $\textrm{d}E=-\epsilon_2\gamma_2\hat{\gamma}(\gamma_2-1)\textrm{d}m_2c^2.$ Substituting $E$ with the formula above, we have
\begin{equation}
\displaystyle{\frac{\textrm{d}\gamma_2}{\textrm{d}m_2}=-\frac{4(\gamma_2^2-1)}
{8(1-\epsilon_2)\gamma_2m_2+3\epsilon_2m_2+3M_{\rm{ej}}}}.
\end{equation}
We have compared our Equation (1) with that of Pe'er (2012). There is almost no difference in the evolution of $\gamma_2$. So this equation can satisfactorily describe the fireball before the density-jump. Most recently, Nava et al. (2013) have proposed a realistic description for the dynamic evolution by including the adiabatic losses in a self-consistent way. Their calculation is slightly more complicated when taking the reverse shock into consideration. We prefer to adopt the simple one here and this would not lead to significant errors on the results.

\subsection{When the Reverse Shock Crossing the Shell}
The interaction of the blast wave with the high-density medium is ascribed to two shocks: a reverse shock that propagates into the hot shell and a forward shock that propagates into the high-density medium. Thus the system consists of four separated regions: (1)unshocked high-density medium, (2) forward-shocked high-density medium, (3) reverse-shocked hot shell, and (4) unshocked hot shell. In this paper, the quantities (Lorentz factor, density, mass, internal energy, pressure) of region ``$i$'' (regions 2, 3 and 4) are denoted by subscripts ``$i$'', respectively. $\gamma_{ij}$ and $\beta_{ij}$ are the relative Lorentz factor and dimensionless velocity of region ``$i$'' measured in the rest frame of region ``$j$''. Similar to the derivation in Section 2.1, we have the total energy of three regions
\begin{equation}
E_{\rm t} = E_2 + E_3 + E_4,
\end{equation}
\begin{equation}
E_i = (\gamma_i-1) m_i c^2 + (1-\epsilon_i) \gamma_i (U_i^{\prime} + P_i^{\prime}),~~~~~~~i=2,3,4.
\end{equation}
With some derivations (see Appendix A), we can get the equation for the evolution of the Lorentz factor of regions 2 and 3 ($\gamma_2=\gamma_3$):
\begin{equation}
\displaystyle{\frac{\textrm{d}\gamma_2}{\textrm{d}m_2}=-\frac{\frac{4}{3}(\gamma_2^2-1)+
f_1\displaystyle\frac{\textrm{d}m_3}{\textrm{d}m_2}+
(1-\epsilon_3)f_2\gamma_2\gamma_{42}(1-\displaystyle\frac{\beta_{42}}{\beta_4})m_3\displaystyle\frac{\textrm{d}\psi_4}{\textrm{d}m_2}}
{\frac{8}{3}(1-\epsilon_2)\gamma_2m_2+\epsilon_2m_2+(1-\epsilon_3)f_3m_3+\epsilon_3m_3}},
\end{equation}
where $f_1$, $f_2$, $f_3$ are functions of other variables (see Appendix A). The evolution equations for $\gamma_4$
and $\psi_4$ (see the definition at Appendix A) can also be obtained together.

\subsection{After the Reverse Shock}
After the reverse shock crosses region 4, only the forward shock is left in the system now. Region 3 becomes the tail of region 2. We can get the hydrodynamical equation of $\gamma_2$ in the way similar to the section above (see Appendix B):
\begin{equation}
\displaystyle{\frac{\textrm{d}\gamma_2}{\textrm{d}m_2}=-\frac{\displaystyle\frac{4}{3}(\gamma_2^2-1)+(1-\epsilon_3)\displaystyle{\left[\frac{4}{3}+\frac{1}{3(1+\psi_3)^2}\right]}\gamma_2m_3\displaystyle\frac{\textrm{d}\psi_3}{\textrm{d}m_2}}
{\displaystyle\frac{8}{3}(1-\epsilon_2)\gamma_2m_2+\epsilon_2m_2+(1-\epsilon_3)\displaystyle{\left[1+\frac{4\psi_3+5}{3(1+\psi_3)}\psi_3\right]}m_3+\epsilon_3m_3}}.
\end{equation}

\section{RADIATION}
\label{sect:radi}
We briefly describe the radiation from the shocked electrons based on the standard
model in this section. Unless special notes, the quantities involving in equations in this section mean the quantities in the shock comoving frame.

In the absence of radiation losses, the energy distribution of shock accelerated electrons behind the shock is usually assumed to be a power-law as $dN_e/d\gamma_e \propto \gamma_e^{-p}$, where $\gamma_e$ is the Lorentz factor of electron and $p$ is the spectrum index. Since the electrons are cooled by synchrotron and inverse Compton (IC) radiation (Rybicki \& Lightman 1979, Sari et al. 1998), the actual electron distribution should be given according to the following cases (Dai et al. 1999):

1. for $\gamma_{e,\rm c}\leq\gamma_{e,\rm m}$,
\begin{equation}
\frac{dN_{e}}{d\gamma_{e}}\propto\left\{\begin{array}{ll}
\gamma_{e}^{-2},&\gamma_{e,\rm c}\leq\gamma_{e}\leq\gamma_{e,\rm m}\\
\gamma_{e}^{-p-1},&\gamma_{e,\rm m}<\gamma_{e}\leq\gamma_{e,\rm
max}\end{array}\right.
\end{equation}

2. for $\gamma_{e,\rm m}<\gamma_{e,\rm c}\leq\gamma_{e,\rm max}$,
\begin{equation}
\frac{dN_{e}}{d\gamma_{e}}\propto\left\{\begin{array}{ll}
\gamma_{e}^{-p},&\gamma_{e,\rm m}\leq\gamma_{e}\leq\gamma_{e,\rm c}\\
\gamma_{e}^{-p-1},&\gamma_{e,\rm c}<\gamma_{e}\leq\gamma_{e,\rm
max}\end{array}\right.
\end{equation}
which are normalized by the total number of the electrons solved from the dynamic equations. The minimum, cooling, and maximum Lorentz factors of electrons are, respectively,
\begin{eqnarray}
\gamma_{e,\rm m}&=&\epsilon_{e}\frac{p-2}{p-1}\frac{m_{\rm
p}}{m_{\rm e}}(\gamma-1),\nonumber\\
\gamma_{e,\rm c}&=&\frac{6\pi m_{\rm e}c}{(1+Y)\sigma_{T}{B}^{2}
(\gamma+\sqrt{\gamma^{2}-1})(t-t_0)},\\
\gamma_{e,\rm max}&=&\sqrt{6\pi q_{\rm e}\over\sigma_{T}B(1+Y)},\nonumber
\end{eqnarray}
where $B$ is the comoving magnetic field strength, $q_{\rm e}$ is the the electric charge of electrons, $m_{\rm p}$ and $m_{\rm e}$ are the mass of proton and electron respectively, $\sigma_{T}$ is the Thomson cross section, $t_0$ is the starting time of each episode in the observer frame and $Y$ is the Compton parameter that is defined as the ratio of the IC power to the synchrotron power. The Compton parameter of an electron with random Lorentz factor $\gamma_e$ is given by $Y(\gamma_e)=(-1+\sqrt{1+4\eta_{\rm rad}\eta_{\rm KN}\epsilon_{e}/\epsilon_{B}})/2$ (He et al. 2009; Fan \& Piran 2006, Wang et al. 2010), where $\eta_{\rm rad}$ is the fraction of the electron's energy that is radiated, $\eta_{\rm KN}$ is the fraction of synchrotron photons below the Klein-Nishina limit, $\epsilon_e$ and $\epsilon_B$ are shock energy equipartition parameters for electrons and magnetic fields respectively.

With the electron distribution determined, the synchrotron radiation flux density can be calculated following previous work (Rybicki \& Lightman 1979; Huang et al. 2000a). The effect of equal-arrival-time surface (EATS; e.g., Waxman 1997; Granot et al. 1999) is considered in the calculations.

\section{CALCULATION RESULTS}
\label{sect:appl}
The equations above can describe the overall evolution of a blast wave encountering a sudden change in density.
In order to explore the difference between our results and previous works, we show the numerical results of dynamics first. We calculate two different cases with density-jump magnitude (denoted as $a$) of 10 and 100 times respectively. We further assume that the blast wave meets the density wall at $\gamma_2=10$ or $\gamma_2=20$. Below, we denote the four different calculations as $\gamma10a10$, $\gamma10a100$, $\gamma20a10$, and $\gamma20a100$
(like Gat et al. 2013). In our calculations, the typical values (e.g., Huang et al. 2000b; Freedman
\& Waxman 2001; Wu et al. 2003) adopted for parameters of the outflow are $E_{K,\rm iso} = 1.0\times10^{53}$ erg, $\theta_j = 0.1$ rad, $p = 2.3$, $\epsilon_e = 0.1$, $\epsilon_B = 0.01$, $\Gamma_0 = 300$, $n_0 = 1.0$ cm$^{-3}$, where $E_{K,\rm iso}$ is the initial isotropic kinetic energy, $\Gamma_0$ and $\theta_j$ are the initial Lorentz factor and half-opening angle of the jet.

Figure 1 shows the evolution of Lorentz factors for different components during the entire time. Each panel in Figure 1 corresponds to one calculation. Before the encounter, the Lorentz factor of the forward shock satisfies the BM solution ($\gamma \propto t^{-3/8}$) quit well. During the reverse shock crossing time, there is a short ``plateau'' for the evolution of $\gamma_2$. This is expected because regions 2 and 3 are now being ``pushed'' by the faster region 4 behind them, $\gamma_2$ will be a constant when an equilibrium between the deceleration by the medium and the acceleration from region 4 (i.e., $\textrm{d}\gamma_2 \simeq 0$) is established. After the reverse shock crossing, the bulk motion of regions 2 and 3 transitions to a trajectory shallower than the BM solution since they are already in the transitional period between the ultrarelativistic and the non-relativistic phases. However, in the non-relativistic phase at fairly late time, the evolution of $\gamma_2$ obeys the Sedov-Von Neumann-Taylor solution $\beta \propto R^{-3/2}$ (Sedov 1959). Comparing the left panels with the right panels in Figure 1, it is within the expectation that the electrons in region 3 are hotter for an earlier density jump, which is more favorable to the re-brightening. The comparison between the upper panels and the lower panels shows that the larger the jump amplitude is, the stronger the reverse shock will be. However, the response in afterglow light curves depends weakly on the the jump amplitude (see below).

It is also convenient to investigate the difference of the hydrodynamics between others' results and ours. Taking the upper left panel for explication, the Lorentz factor ratio at the encounter time $\gamma_4/\gamma_2 (t=t_{enc})$ is $\sim 2.2$ from our result, while the values calculated from other analytical equations are $\sim3.2$ (Dai \& Lu 2002) and $\sim 1.7$ (NG: Nakar \& Granot 2007) respectively. In contrast, the lower $\gamma_2$ and higher $\gamma_{th,3}$ predicted by Dai \& Lu (2002) would lead to an overestimate of the flux emitted from hot electrons during this episode. The little discrepancy between NG and ours is caused by the different approximations used when dealing with the conservative equations and it does not lead qualitatively change to our conclusions below.

Before calculating the radiation, it is notable to look at the number density of hot electrons in region 3. It is crucial to take the radial structure of blast wave into account according to the discussion in van Eerten et al. (2009). In fact, van Eerten and his collaborators emphasized the difference in density of region 3 between the simulation and the analytical results is another significant factor leading to the contradiction of their conclusions. So it needs to be prudent when dealing with this density in our work. Although we have made the thin shell assumption when deriving the hydrodynamic equations, it is easy for us to calculate the ``real'' (volume-averaged) density of region 3 here in our model. Using the velocities of forward shock and the reverse shock, we can obtain the real co-moving width of region 3 and the corresponding density (see Equation A24), which will replace the density derived from the rough jump conditions. Likewise, the real density in our model is much lower than that predicted by jump conditions. Figure 2 depicts the comoving density profile during the encounter at the observer time $t = 40900 $ s ($z = 1.0$) for the upper left panel in Figure 1, the real density of region 3 is $\sim 3$ times lower than analytical one. Note that our density does not satisfy the shock jump condition, it is due to the simplified homogeneous assumption adopted by us. In reality, the hydrodynamic quantities should satisfy the shock jump condition at the shock and have gradients in the shocked region. This simplified assumption will not lead to severe error.

Afterglow light curves can be obtained by considering the radiation process described in Section 3 (assuming a redshift $z = 1$). Figure 3 shows the light curves in X-ray and optical band for the four calculations in Figure 1. The emission from region 3 is found to be less than or comparable to emission from high latitude early shell, no notable bumps emerge after the density-jump in these four cases. In each panel, the total light curve smoothly transits to a steep decay phase after the encounter. In fact, the total emission after the encounter is mainly determined by the curvature effect (Fenimore et al. 1996) --- after the reverse shock crossing time, no more fresh hot electrons are injected into region 3. According to the scaling relation of curvature effect $F_{\nu} \propto t^{-(2+\beta)}$ ($\beta$ is the spectrum index), the temporal indices are then $-2.65$ (slow cooling) or $-3.15$ (fast cooling) for $p = 2.3$. The slopes of the light curves in Figure 3 are just within this range. The lower component, emission from region 2 first raises steeply when $\gamma_2$ stays at the plateau phase, then decays sharply due to the decrease of $\gamma_2$ and comoving density of region 2. And at last it turns to a normal decay close to the BM solution. The character of the steep decay after the encounter in our model is not seen in the simulation results of Gat et al. (2013). This is the manifestation of the deviation caused by the homogenous thin shell assumption (i.e., sharp edge of each region) in our work. With the radial structure of blast wave considered in the full hydrodynamics simulation, the result of a shallower and smoother light curve is likely expected.

If the Lorentz factor just before the jump is much larger (e.g., $\gamma_2 > 20$), a plateau would emerge in our model. However, this condition is often not satisfied since we concern about the late afterglow ($t > 10^4$ s). The contribution of reverse shock component is mild and slightly depends on the jump amplitude from the comparison between the upper and lower panels.

We do the same calculations for the cases in which the blast wave travels in a stellar wind environment before the encounter and enters into a homogeneous ISM after the encounter. The visible response to the density-jump as showed in Figure 4 is slighter compared with the entire ISM case. In the wind environment, the prior temporal index of light curve in the left panel is $(1-3p)/4 \sim -1.5$ (Zhang \& M{\'e}sz{\'a}ros 2004) and is steeper than that of the ISM cases. Thus the smoother transition from the prior light curve to the phase after the encounter seems reasonable. Light curves for blast wave encountering a density change of various magnitudes at a late time ($\gamma_4=3$) are also presented in Figure 5. Figure 5 depicts that there are no observable bumps for various density jumps in this case too. For the blast waves traveling from a stellar wind to the ISM (right panel), the shallow behavior of optical afterglows are similar to those in Gat et al. (2013). According to the context above, the lower the encounter Lorentz factor is, the weaker the emission flux of region 3 is. Thus the flux decay during the encounter is naturally explained by the domination of tail emission of region 4 in our model.

\section{DISCUSSION AND CONCLUSIONS}
\label{sect:disc}
A set of extended differential equations for relativistic outflows encountering a density-jump medium are derived in this paper. Using these, we have obtained a more accurate dynamics (Figure 1) and calculated an appropriate density of region 3 (Figure 2) in this issue. The extended equations in this article can also be applied to study the collision of two homogeneous shells after minor modifications. According to our numerical results, no high-amplitude re-brightening would emerge when the hot electrons in region 3 begin to contribute to the emission flux after encountering the density wall. Van Eerten et al. (2009) and Gat et al. (2013) concluded that sudden transition in circumburst density is very unlikely to be the cause of the bumps using adaptive mesh relativistic hydrodynamic codes. Taking a homogeneous slab and ignoring the radial structure of blast wave in previous analytical result may lead to an overestimation. Although there are still some discrepancies between the simulation results and the results from our semi-analytic model, probably due to the thin shell assumption imprinted in this work, our results disfavor the density-jump origin for bumps on afterglows from another perspective.

Generally speaking, the density-jump scenario cannot explain the observed late re-brightenings in many GRBs. Other scenarios involving late refreshed shocks or late activities of the central engine may explain these re-brightenings (Margutti et al. 2011). The late internal shock model can naturally explain the X-ray flares with sharp profiles (Falcone et al. 2007). Recent works also show that the central engine (a black hole) may be re-activated after the initial burst. The fallback of material onto the central black hole after the collapse could last for a long time (Woosley 1993; MacFadyen et al. 2001; Perna et al. 2014) and lead to late central engine activities (Perna et al. 2006; Kumar et al. 2008a,b). The observational evidence for this process is highlighted in Wu et al. (2013) and Yu et al. (2013). The two-component jet model scenario may also be consistent with the gradual re-brightening of some GRB afterglows (Huang et al. 2004; Peng et al. 2005; Filgas et al. 2011). Thus, it still remains to be answered which one or how many of these scenarios (refreshed shocks, late activities of central engines, two-component jets) plays the key role in the observation sample. The spectral evolution during the bump may help to test the models for individual GRBs (Nardini et al. 2011b). Radio observations of afterglows are also valuable for theoretical modeling (Moin et al. 2013). It is interesting to see that some bumps may be related with the central engine activities, which makes it helpful for studying the properties of progenitor stars.

\acknowledgments
We thank the anonymous referee for helpful comments and suggestions that lead to an overall improvement of this study. This work was supported by the National Basic Research Program of China (973 Program, Grant No. 2014CB845800, and 2013CB834900) and the National Natural Science Foundation of China (Grant No. 11033002 and 11322328). X. F. Wu acknowledges support by the One-Hundred-Talents Program, the Youth Innovation Promotion Association, and the Strategic Priority Research Program ¡°The Emergence of Cosmological Structures¡± of the Chinese Academy of Sciences (Grant No. XDB09000000). Liang Li acknowledges support by the Swedish National Space Board, and the Erasmus Mundus Joint Doctorate Program by Grant Number 2011-1640 from the EACEA of the European Commission.

\appendix
\section{EQUATIONS FOR THE SECOND EPISODE}
We derive the differential equations for the dynamics of the system when the reverse shock exists. The total energy of regions 2, 3 and 4 are
\begin{equation}
E_{\rm t} = E_2 + E_3 + E_4,
\end{equation}
\begin{equation}
E_i = (\gamma_i-1) m_i c^2 + (1-\epsilon_i) \gamma_i (U_i^{\prime} + P_i^{\prime}),~~~~~~~i=2,3,4,
\end{equation}
where
\begin{eqnarray}
U_2^{\prime}&=&\psi_2 m_2 c^2=(\gamma_2-1)m_2 c^2, \nonumber \\
U_3^{\prime}&=&\psi_3 m_3 c^2=(\gamma_{34}-1) m_3 c^2+\gamma_{34}(U_3^{\prime}+P_3^{\prime})=\left[\gamma_{34}\frac{4\psi_4+5}{3(1+\psi_4)}\psi_4+(\gamma_{34}-1)\right]m_3 c^2, \\
U_4^{\prime}&=&\psi_4 m_4 c^2, \nonumber \\
P_{\rm i}^{\prime}&=&\frac{\psi_{\rm{i}}+2}{3(1+\psi_{\rm{i}})}\psi_{\rm{i}}m_{\rm{i}}c^2.\nonumber
\end{eqnarray}
The parameter $\psi_4$ is time dependent and the expression for it will be obtained below. At the time when the reverse shock emerges ($R = R_0$), we can get the initial value of $\psi_4$: $\psi_{4,0} = \frac{U_{4,0}^{\prime}}{m_{4,0} c^2}=\gamma_{4,0}-1$, where $\gamma_{4,0}$ equals to the value of $\gamma_2$ at $R_0$. Meanwhile, the total mass of regions 3 and 4 is $m_{4,0}=m_3+m_4$. Substituting Equations (A2) and (A3) into Equation (A1), the total energy of the system is
\begin{eqnarray}
E_{\rm t}&=&\frac{4}{3}(\gamma_2^2-1)m_2
c^2-\epsilon_2\frac{4\gamma_2+1}{3}(\gamma_2-1)m_2
c^2 \nonumber \\
         &&+(\gamma_3-1)m_3c^2+(1-\epsilon_3)\gamma_3\frac{4\psi_3+5}{3(1+\psi_3)}\psi_3 m_3c^2
         \nonumber \\
         &&+(\gamma_4-1)m_4 c^2+\gamma_4\frac{4\psi_4+5}{3(1+\psi_4)}\psi_4 m_4
         c^2, \,\,\,
\end{eqnarray}
where we have assumed that $\gamma_2=\gamma_3$ and $\epsilon_4=0$. The unshocked portion of region 4 will not lose its energy, although conversion of the thermal energy to the bulk kinetic energy may happen. Thus we have $\displaystyle\gamma_4\left[1+\frac{4\psi_4+5}{3(1+\psi_4)}\psi_4\right]=\gamma_{4,0}\left[1+\frac{4\psi_{4,0}+5}{3(1+\psi_{4,0})}\psi_{4,0}\right]
=\frac{4\gamma_{4,0}^2-1}{3}$, or
\begin{equation}
\textrm{d}\gamma_4=-\frac{4(1+\psi_4)^2+1}{(2\psi_4+1)(2\psi_4+3)}\frac{\gamma_4}{1+\psi_4}\textrm{d}\psi_4.
\end{equation}
Let us have a look at the evolution of $\psi$ (represents $\psi_4$) due to adiabatic expansion. We discuss it in the co-moving frame. For a system with the mass $m$, pressure $p^{\prime}$ and volume $V^{\prime}$, the thermal Lorentz factor is $\gamma_{th}=1+\psi$. We take the adiabatic index as
$\hat{\gamma}_{th}\simeq\displaystyle\frac{4\gamma_{th}+1}{3\gamma_{th}}$,
so the pressure is
\begin{eqnarray}
p^{\prime}&=&(\hat{\gamma}_{th}-1)e^{\prime}=(\hat{\gamma}_{th}-1)(\gamma_{th}-1)\rho^{\prime}c^2
\nonumber \\
          &=&\displaystyle\frac{\gamma_{th}^2-1}{3\gamma_{th}}\rho^{\prime}c^2=\frac{\psi(2+\psi)}{3(1+\psi)}\rho^{\prime}c^2.
\end{eqnarray}
The equation of adiabatic expansion is
\begin{equation}
mc^2\textrm{d}\psi=-p^{\prime}\textrm{d}V^{\prime}.
\end{equation}
The above equation leads to the solution of
\begin{equation}
\frac{(1+\psi)^2-1}{(1+\psi_0)^2-1}=(\frac{\rho^{\prime}}{\rho_0^{\prime}})^{\frac{2}{3}}.
\end{equation}
So the evolution of $\psi_4$ can be written as
\begin{eqnarray}
\frac{\textrm{d}\psi_4}{\textrm{d}m_2}&=&\frac{\textrm{d}\psi_4}{\textrm{d}\ln{\rho_4^{\prime}}}
\frac{\textrm{d}\ln{\rho_4^{\prime}}}{\textrm{d}\ln{R}}\frac{\textrm{d}\ln{R}}{\textrm{d}m_2}
 \nonumber\\
 &=&\frac{\psi_4(2+\psi_4)}{3(1+\psi_4)}\frac{\textrm{d}\ln{\rho_4^{\prime}}}{\textrm{d}\ln{R}}
    \frac{\textrm{d}R}{R\textrm{d}m_2},
\end{eqnarray}
where
\begin{eqnarray}
\frac{\textrm{d}\ln{\rho_4^{\prime}}}{\textrm{d}\ln{R}}
&=&\frac{\partial\ln{\rho_4^{\prime}}}{\partial\ln{R}}|_{\theta_4,\gamma_4}
+\frac{\partial\ln{\rho_4^{\prime}}}{\partial\ln{\theta_4}}|_{\gamma_4,R}\frac{\textrm{d}\ln{\theta_4}}{\textrm{d}\ln{R}}
+\frac{\partial\ln{\rho_4^{\prime}}}{\partial\ln{\gamma_4}}|_{R,\theta_4}\frac{\textrm{d}\ln{\gamma_4}}{\textrm{d}\ln{R}}
\nonumber \\
&=&\frac{\partial\ln{\rho_4^{\prime}}}{\partial\ln{R}}|_{\theta_4,\gamma_4}-\frac{\sin{\theta_4}}{1-\cos{\theta_4}}\frac{c_{s,4}}{\beta_4\gamma_4c}
+\frac{\partial\ln{\rho_4^{\prime}}}{\partial\ln{\gamma_4}}|_{R,\theta_4}\frac{\textrm{d}\ln{\gamma_4}}{\textrm{d}\ln{R}}.
\end{eqnarray}
The second equality has included the lateral expansion of the shell (detailed expression will be showed in Equation (A33)). Equation (A5) and Equations (A9)--(A10) lead to
\begin{equation}
\frac{\textrm{d}\psi_4}{\textrm{d}m_2}=\frac{(1+\psi_4)\displaystyle\left[\frac{\partial\ln{\rho_4^{\prime}}}{\partial\ln{R}}|_{\theta_4,\gamma_4}
+\frac{\partial\ln{\rho_4^{\prime}}}{\partial\ln{\theta_4}}|_{\gamma_{4},R}\frac{\textrm{d}\ln{\theta_4}}{\textrm{d}\ln{R}}\right]\frac{\textrm{d}\ln{R}}{\textrm{d}m_2}}
{\displaystyle{\frac{3(1+\psi_4)^2}{\psi_4(2+\psi_4)}+\frac{\partial\ln{\rho_4^{\prime}}}{\partial\ln{\gamma_4}}|_{R,\theta_4}}}.
\end{equation}
We can also get
\begin{eqnarray}
\frac{\textrm{d}\gamma_4}{\textrm{d}R}
&=&-\frac{\displaystyle{\frac{\partial\ln{\rho_4^{\prime}}}{\partial\ln{R}}|_{\theta_4,\gamma_4}
+\frac{\partial\ln{\rho_4^{\prime}}}{\partial\ln{\theta_4}}|_{\gamma_{4},R}\frac{\textrm{d}\ln{\theta_4}}{\textrm{d}\ln{R}}}}
{\displaystyle{\frac{3(1+\psi_4)^2}{\psi_4(2+\psi_4)}\frac{(2\psi_4+1)(2\psi_4+3)}{4(1+\psi_4)^2+1}+\frac{\partial\ln{\rho_4^{\prime}}}{\partial\ln{\gamma_4}}|_{R,\theta_4}}}
\frac{\gamma_4}{R}
\nonumber \\
&=&\frac{\displaystyle{\frac{\sin{\theta_4}}{1-\cos{\theta_4}}\frac{c_{s,4}}{\beta_4\gamma_4c}-\frac{\partial\ln{\rho_4^{\prime}}}{\partial\ln{R}}|_{\theta_4,\gamma_4}}}
{\displaystyle{\frac{3(1+\psi_4)^2}{\psi_4(2+\psi_4)}\frac{(2\psi_4+1)(2\psi_4+3)}{4(1+\psi_4)^2+1}+\frac{\partial\ln{\rho_4^{\prime}}}{\partial\ln{\gamma_4}}|_{R,\theta_4}}}
\frac{\gamma_4}{R}.
\end{eqnarray}
On the other hand, the radiative energy is
\begin{equation}
\textrm{d}E_{\rm t}=-\frac{1}{3}\epsilon_2(4\gamma_2+1)(\gamma_2-1)\textrm{d}m_2
c^2-\epsilon_3\gamma_3\frac{4\psi_3+5}{3(1+\psi_3)}\psi_3\textrm{d}m_3
c^2.
\end{equation}
Combining Equations (A1)--(A4) and (A13), and set $\epsilon_4=0$, we get
\begin{eqnarray}
& &
\frac{4}{3}(\gamma_2^2-1)\textrm{d}m_2+\frac{8}{3}\gamma_2m_2\textrm{d}\gamma_2+\gamma_2\left[1+\frac{4\psi_3+5}{3(1+\psi_3)}\psi_3\right]\textrm{d}m_3
 +m_3\textrm{d}\gamma_2 \nonumber \\
& &
+(1-\epsilon_3)\frac{4\psi_3+5}{3(1+\psi_3)}\psi_3m_3\textrm{d}\gamma_2+(1-\epsilon_3)\left[\frac{4}{3}+\frac{1}{3(1+\psi_3)^2}\right]\gamma_2m_3\textrm{d}\psi_3
=\nonumber \\
&
&\gamma_4\left[1+\frac{4\psi_4+5}{3(1+\psi_4)}\psi_4\right]\textrm{d}m_3+\epsilon_2\frac{8\gamma_2-3}{3}
m_2\textrm{d}\gamma_2.
\end{eqnarray}
Taking $\gamma_{34}=(1-\beta_3\beta_4)\gamma_3\gamma_4$, $\gamma_2=\gamma_3$, we have
\begin{equation}
\;\;\;\;\;\;\textrm{d}\gamma_{34}=(1-\frac{\beta_4}{\beta_2})\gamma_4\textrm{d}\gamma_2
+(1-\frac{\beta_2}{\beta_4})\gamma_2\textrm{d}\gamma_4=
\frac{\beta_{24}\gamma_{24}}{\beta_2\gamma_2}\textrm{d}\gamma_2+
\frac{\beta_{42}\gamma_{42}}{\beta_4\gamma_4}\textrm{d}\gamma_4.
\end{equation}
For $m_3$, since
\begin{eqnarray}
& &\textrm{d}m_2=2\pi(1-\cos{\theta_2}) R^2\rho_1
\textrm{d}R=2\pi(1-\cos{\theta_2}) R^2\rho_1\beta_2
c\textrm{d}t_b, \nonumber \\
& &
\textrm{d}m_3=2\pi(1-\cos{\theta_3})R^2\rho_4^{\prime}\gamma_4(\beta_4-\beta_3)c\textrm{d}t_b,
\end{eqnarray}
we have
\begin{equation}
\textrm{d}m_3=\displaystyle{(\frac{\beta_4}{\beta_2}-1)\frac{\rho_4^{\prime}}{\rho_1}\frac{1-\cos{\theta_3}}{1-\cos{\theta_2}}
\gamma_4\textrm{d}m_2},
\end{equation}
where $\textrm{d}t_b=\gamma_2(\gamma_2+\sqrt{\gamma_2^2-1})\textrm{d}t$ is measured in the burst's frame while $\textrm{d}t$ is the difference in arrival times (we neglect the ($1+z$) term here). $\rho_1$ is the environment density at $R>R_0$, and
\begin{equation}
\rho_4^{\prime}=\rho_{4,0}^{\prime}\frac{\gamma_4}{\gamma_{4,0}}\frac{R_0^3}{R^3}\frac{1-\cos{\theta_{4,0}}}{1-\cos{\theta_4}}
\end{equation}
is the co-moving density of region 4, $\rho_{4,0}^{\prime}$ and $\theta_{4,0}$ are the parameters at $R_0$. Inserting Equations (A3) and (A15) into (A14), we have the evolution of $\gamma_2$ during the reverse shock crossing
\begin{equation}
\displaystyle{\frac{\textrm{d}\gamma_2}{\textrm{d}m_2}=-\frac{\frac{4}{3}(\gamma_2^2-1)+f_1\displaystyle\frac{\textrm{d}m_3}{\textrm{d}m_2}+(1-\epsilon_3)f_2\gamma_2\gamma_{42}(1-\displaystyle\frac{\beta_{42}}{\beta_4})m_3
\displaystyle\frac{\textrm{d}\psi_4}{\textrm{d}m_2}}
{\frac{8}{3}(1-\epsilon_2)\gamma_2m_2+\epsilon_2m_2+(1-\epsilon_3)f_3m_3+\epsilon_3m_3}},
\end{equation}
where
\begin{equation}
f_1=\gamma_2\left[1+\frac{4\psi_3+5}{3(1+\psi_3)}\psi_3\right]-\gamma_4\left[1+\frac{4\psi_4+5}{3(1+\psi_4)}\psi_4\right],
\end{equation}
\begin{equation}
f_2=\left[\frac{4}{3}+\frac{1}{3(1+\psi_3)^2}\right]\left[\frac{4}{3}+\frac{1}{3(1+\psi_4)^2}\right],
\end{equation}
\begin{equation}
f_3=1+\frac{4\psi_3+5}{3(1+\psi_3)}\psi_3-\left[\frac{4}{3}+\frac{1}{3(1+\psi_3)^2}\right]\left[1+\frac{4\psi_4+5}{3(1+\psi_4)}\psi_4\right]
\frac{\beta_{42}}{\beta_2}\gamma_{42}.
\end{equation}
Although we have made the thin shell assumption, we would calculate the co-moving width of region 3
\begin{equation}
\textrm{d}\Delta_3^{\prime}=\frac{\gamma_3(\beta_4-\beta_3)c\textrm{d}t_b}{\frac{\gamma_3 \rho_3^{\prime}}{\gamma_4 \rho_4^{\prime}}-1}
\end{equation}
and its corresponding ``real'' (volume-averaged) density
\begin{equation}
\rho_3^{\prime}=\displaystyle{\frac{m_3}{2\pi (1-\cos{\theta_3}) R^2 \Delta_3^{\prime}}}.
\end{equation}
Van Eerten (2009) pointed out that the radial structure of blast wave is relevant with the density-jump issue.
The real density here is also a good approach to the actual case to some extent. In Equation (A23), we take $\rho_3^{\prime} = (4\gamma_{34}+3)\rho_4^{\prime}$ (jump condition) for the calculation of each small increment of $\Delta_3^{\prime}$.

We extend the jump condition to either cold (Sari \& Piran 1995) or hot shell after the reverse shock emerges by
\begin{equation}
(1+\psi_4)(\frac{\beta_4}{\beta_2}-1)^2\frac{\rho_4^{\prime}}{\rho_1}\gamma_4^2=1,
\end{equation}
with the criterion for the formation of the reverse shock, i.e. $\beta_{42}c>c_{s,4}$ (sound speed of region 4 in the co-moving frame).

If the shell is hot (corresponding to the density-jump case), $1+\psi_4=\gamma_4$, $\rho_4^{\prime}=4\gamma_4\rho_0$, we get
\begin{equation}
\beta_{42}=\frac{\sqrt{\Re}}{2+\sqrt{\Re}}\beta_4, \;\;\;\;
\gamma_{42}=\frac{2+\sqrt{\Re}}{\sqrt{4(1+\sqrt{\Re})\gamma_2^2+\Re}}\gamma_4,
\end{equation}
where
$\Re\equiv\displaystyle\frac{\rho_1}{\rho_0}$, with $\rho_0\propto R^{-k}$($k=0$ for ISM; $k=2$ for wind) at $R\leq R_0$. In relativistic stage, $\gamma_4\gg1$, $c_{s,4}=c/\sqrt{3}$, the criterion for the formation of reverse shock is $\Re>(\sqrt{3}+1)^2\simeq 7.5$. The reverse shock can be relativistic ($\gamma_{42}\geq2$) if $\sqrt{\Re}\geq\displaystyle\frac{6+4\sqrt{3}\beta_4}{4\beta_4^2-3}$
or $\Re\geq168$ ($\gamma_4\gg2$). The analysis here is consistent with the results in Dai \& Lu (2002). In the non-relativistic stage, $\beta_4\ll1$,
$c_{s,4}=\displaystyle\frac{\sqrt{5}}{3}\beta_4 c$, the criterion for the formation of reverse shock becomes
$\Re>(35+15\sqrt{5})/2=34.3$.

If the shell is cold, $\psi_4=0$, we have
\begin{equation}
\beta_{42}=\frac{\gamma_4}{\gamma_4+\sqrt{f}}\beta_4, \;\;\;\;
\gamma_{42}=\frac{\gamma_4+\sqrt{f}}{\sqrt{1+f+2\gamma_4\sqrt{f}}},
\end{equation}
where $f\equiv\displaystyle\frac{\rho_4^{\prime}}{\rho_1}$.
Since the initial shell is cold ($c_{s,4}=0$), the reverse shock can always be developed. The reverse shock can be relativistic ($\gamma_{42}\geq2$) if $\sqrt{f}\leq\displaystyle\frac{2\sqrt{3}\beta_4-3}{3}\gamma_4$,
or $f\leq\gamma_4^2/(3+2\sqrt{3})^2\simeq\gamma_4^2/42$.

However, the observed time of region 4 is shorter than that of regions 3 and 2, this is because of the difference of their bulk Lorentz factor, i.e. $\gamma_4>\gamma_3=\gamma_2$, while the radial increment is the same. The relation between the observed times is
\begin{equation}
\textrm{d}t_4=\displaystyle\frac{\beta_2\gamma_2(\gamma_2+\sqrt{\gamma_2^2-1})}
{\beta_4\gamma_4(\gamma_4+\sqrt{\gamma_4^2-1})}\textrm{d}t.
\end{equation}
We still need other four equations to complete the hydrodynamics, i.e.
\begin{equation}
\displaystyle\frac{\textrm{d}R}{\textrm{d}t}=\beta_2c\gamma_2(\gamma_2+\sqrt{\gamma_2^2-1}),
\end{equation}
\begin{equation}
\displaystyle\frac{\textrm{d}\theta_2}{\textrm{d}t}=\displaystyle\frac{c_{s,2}(\gamma_2+\sqrt{\gamma_2^2-1})}{R},
\end{equation}
\begin{equation}
\displaystyle\frac{\textrm{d}\theta_3}{\textrm{d}t}=\displaystyle\frac{c_{s,3}(\gamma_2+\sqrt{\gamma_2^2-1})}{R},
\end{equation}
\begin{equation}
\displaystyle\frac{\textrm{d}\theta_4}{\textrm{d}t}=\displaystyle\frac{c_{s,4}}{R}
\displaystyle\frac{\beta_2\gamma_2(\gamma_2+\sqrt{\gamma_2^2-1})}{\beta_4\gamma_4},
\end{equation}
or
\begin{equation}
\frac{\textrm{d}\theta_{i}}{\textrm{d}R}=\frac{c_{s,i}}{\beta_{i}\gamma_{i}c}\frac{1}{R},
\end{equation}
where the sound speeds in the co-moving frame are
\begin{equation}
c_{s,i}^2=\hat{\gamma}_{th,i}(\hat{\gamma}_{th,i}-1)(\gamma_{th,i}-1)
\displaystyle\frac{1}{1+\hat{\gamma}_{th,i}(\gamma_{th,i}-1)}c^2,
\end{equation}
in which $\gamma_{th,2}=\gamma_2$ for region 2, $\gamma_{th,3}=\gamma_{42}(1+\psi_4)$ for region 3 and $\gamma_{th,4}=1+\psi_4$ for region 4.

\section{THE EQUATIONS FOR THE THIRD EPISODE}
We derive the differential equations for the dynamics of the system after the reverse shock crossing time $t_{across}$. Region 4 vanishes now and the total energy of the system is
\begin{equation}
E_{\rm t}=E_{\rm{2}}+E_{\rm{3}},
\end{equation}
with
\begin{equation}
E_{\rm{2}}=\frac{4}{3}(\gamma_2^2-1)m_2
c^2-\epsilon_2\frac{4\gamma_2+1}{3}(\gamma_2-1)m_2c^2,
\end{equation}
\begin{equation}
E_{\rm{3}}=(\gamma_2-1)m_3
c^2+(1-\epsilon_3)\gamma_2\frac{4\psi_3+5}{3(1+\psi_3)}\psi_3m_3c^2,
\end{equation}
where $m_3=m_{4,0}=$const,
$\psi_{3,0}=\gamma_{34}\left[1+\displaystyle\frac{4\psi_4+5}{3(1+\psi_4)}\psi_4\right]-1|_{t=t_{across}}$.
The radiative energy is
\begin{equation}
\textrm{d}E_{\rm t}=-\frac{1}{3}\epsilon_2(4\gamma_2+1)(\gamma_2-1)\textrm{d}m_2
c^2.
\end{equation}
The combination of the above four equations leads to
\begin{eqnarray}
& &
\frac{4}{3}(\gamma_2^2-1)\textrm{d}m_2+\frac{8}{3}\gamma_2m_2\textrm{d}\gamma_2
+m_3\textrm{d}\gamma_2 \nonumber \\
& &
+(1-\epsilon_3)\frac{4\psi_3+5}{3(1+\psi_3)}\psi_3m_3\textrm{d}\gamma_2+(1-\epsilon_3)\left[\frac{4}{3}+\frac{1}{3(1+\psi_3)^2}\right]\gamma_2m_3\textrm{d}\psi_3
\nonumber \\
& &=\frac{1}{3}\epsilon_2(8\gamma_2-3)m_2\textrm{d}\gamma_2,
\end{eqnarray}
thus the hydrodynamic equation is
\begin{equation}
\displaystyle{\frac{\textrm{d}\gamma_2}{\textrm{d}m_2}=-\frac{\displaystyle\frac{4}{3}(\gamma_2^2-1)+(1-\epsilon_3)\displaystyle{\left[\frac{4}{3}+\frac{1}{3(1+\psi_3)^2}\right]}\gamma_2m_3\displaystyle\frac{\textrm{d}\psi_3}{\textrm{d}m_2}}
{\displaystyle\frac{8}{3}(1-\epsilon_2)\gamma_2m_2+\epsilon_2m_2+(1-\epsilon_3)\displaystyle{\left[1+\frac{4\psi_3+5}{3(1+\psi_3)}\psi_3\right]}m_3+\epsilon_3m_3}}.
\end{equation}
For simplicity, we set $\epsilon_3=0$, and the final result is
\begin{equation}
\displaystyle{\frac{\textrm{d}\gamma_2}{\textrm{d}m_2}=-\frac{\displaystyle\frac{4}{3}(\gamma_2^2-1)+\displaystyle{\left[\frac{4}{3}+\frac{1}{3(1+\psi_3)^2}\right]}\gamma_2m_3\displaystyle\frac{\textrm{d}\psi_3}{\textrm{d}m_2}}
{\displaystyle\frac{8}{3}(1-\epsilon_2)\gamma_2m_2+\epsilon_2m_2+\displaystyle{\left[1+\frac{4\psi_3+5}{3(1+\psi_3)}\psi_3\right]}m_3}}.
\end{equation}
This result is consistent with the generic model of Huang et al. (1999) if $\psi_3=0$.

The evolution of $\psi_3$ is model dependent. Here we give two scenarios, one is through the work done by region 3 to region 2, another is considered by the adiabatic expansion of region 3. In both scenarios the decrease of the adiabatic thermal energy of region 3 can postpone the deceleration of the bulk motion of both region 3 and region 2. Here, we only discuss the adiabatic expansion scenario.

The evolution of $\psi_3$ also follows the adiabatic expansion of Equation (A9), and
\begin{equation}
\rho_3^{\prime}=\rho_{3,a}^{\prime}\frac{\gamma_2}{\gamma_{2,a}}\frac{R_a^3}{R^3}\frac{1-\cos{\theta_{3,a}}}{1-\cos{\theta_3}},
\end{equation}
in which $\rho_{3,a}^{\prime}\simeq 4\gamma_{42}\rho_4^{\prime}|_{t_{across}}$, $\gamma_{2,a}$ and
$R_a$ are the parameters when the reverse shock just crosses region 4. So
\begin{eqnarray}
\frac{\textrm{d}\ln{\rho_3^{\prime}}}{\textrm{d}\ln{R}}
&=&\frac{\partial\ln{\rho_3^{\prime}}}{\partial\ln{R}}|_{\theta_3,\gamma_2}
+\frac{\partial\ln{\rho_3^{\prime}}}{\partial\ln{\theta_3}}|_{\gamma_2,R}\frac{\textrm{d}\ln{\theta_3}}{\textrm{d}\ln{R}}
+\frac{\partial\ln{\rho_3^{\prime}}}{\partial\ln{\gamma_2}}|_{R,\theta_3}\frac{\textrm{d}\ln{\gamma_2}}{\textrm{d}\ln{R}}
\nonumber \\
&=&-3-\frac{\sin{\theta_3}}{1-\cos{\theta_3}}\frac{c_{s,3}}{\beta_2\gamma_2c}+\frac{\textrm{d}\ln{\gamma_2}}{\textrm{d}\ln{R}}.
\end{eqnarray}
We thus have
\begin{equation}
\frac{\textrm{d}\psi_3}{\textrm{d}m_2}
=\frac{\psi_3(2+\psi_3)}{3(1+\psi_3)}\frac{\textrm{d}\ln{\rho_3^{\prime}}}{\textrm{d}\ln{R}}
    \frac{\textrm{d}R}{R\textrm{d}m_2},
\end{equation}
and
\begin{equation}
\frac{\textrm{d}\gamma_2}{\textrm{d}m_2}=-\frac{\displaystyle\frac{4}{3}(\gamma_2^2-1)-\gamma_2m_3\displaystyle\frac{\psi_3(2+\psi_3)}{3(1+\psi_3)}
\left[\frac{4}{3}+\frac{1}{3(1+\psi_3)^2}\right]\left[3+\frac{\sin{\theta_3}}{1-\cos{\theta_3}}\frac{c_{s,3}}{\beta_2\gamma_2c
}\right]\displaystyle\frac{\textrm{d}R}{R\textrm{d}m_2}}
{\displaystyle\frac{8}{3}(1-\epsilon_2)\gamma_2m_2+\epsilon_2m_2+\displaystyle\frac{(2\psi_3^2+4\psi_3+1)(8\psi_3^2+16\psi_3+9)}{9(1+\psi_3)^3}m_3}.
\end{equation}

\begin{figure}
   \begin{center}
   \includegraphics[scale=0.6]{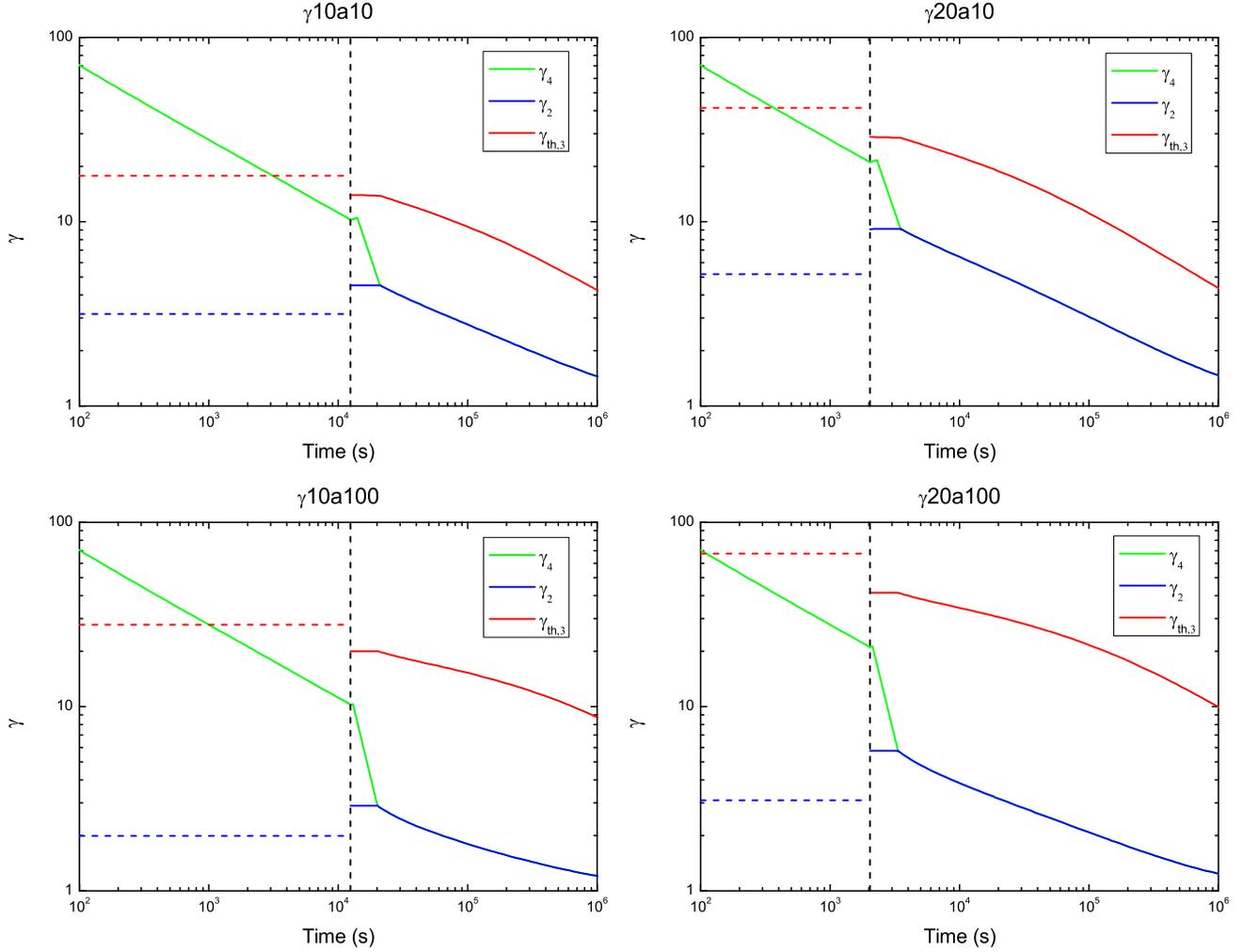}
   \caption{Evolution of Lorentz factors for different components. The encountering time is marked by the vertical dashed line. The green solid lines represent the bulk Lorentz factor of the forward shock before encountering the density-jump and the bulk Lorentz factor of region 4 during the encountering, both denoted by $\gamma_4$. The blue solid lines and red solid lines are the bulk Lorentz factors of regions 2, 3 ($\gamma_2 = \gamma_3$) and the thermal Lorentz factor ($\gamma_{th,3}$) of the baryons in region 3 after the encountering respectively. The horizontal dashed lines represent the corresponding Lorentz factors (remain to be constant during the reverse shock crossing time) given by Dai \& Lu (2002). The four panels correspond to the four cases with different initial conditions defined in Section 4. Note that $\gamma_4$ (see the green solid lines) during the encounter is almost constant (slowly increasing due to adiabatic expansion), which is showed as a ``plateau'' that lasts much shorter than that of regions 2, 3. The shortness of this ``plateau'' of this region is just due to the different transformation formula between the burst's frame time and the apparent time in the observer frame (see Equation A28).}
   \label{Fig:plot1}
   \end{center}
\end{figure}

\begin{figure}
   \begin{center}
   \includegraphics[scale=0.6]{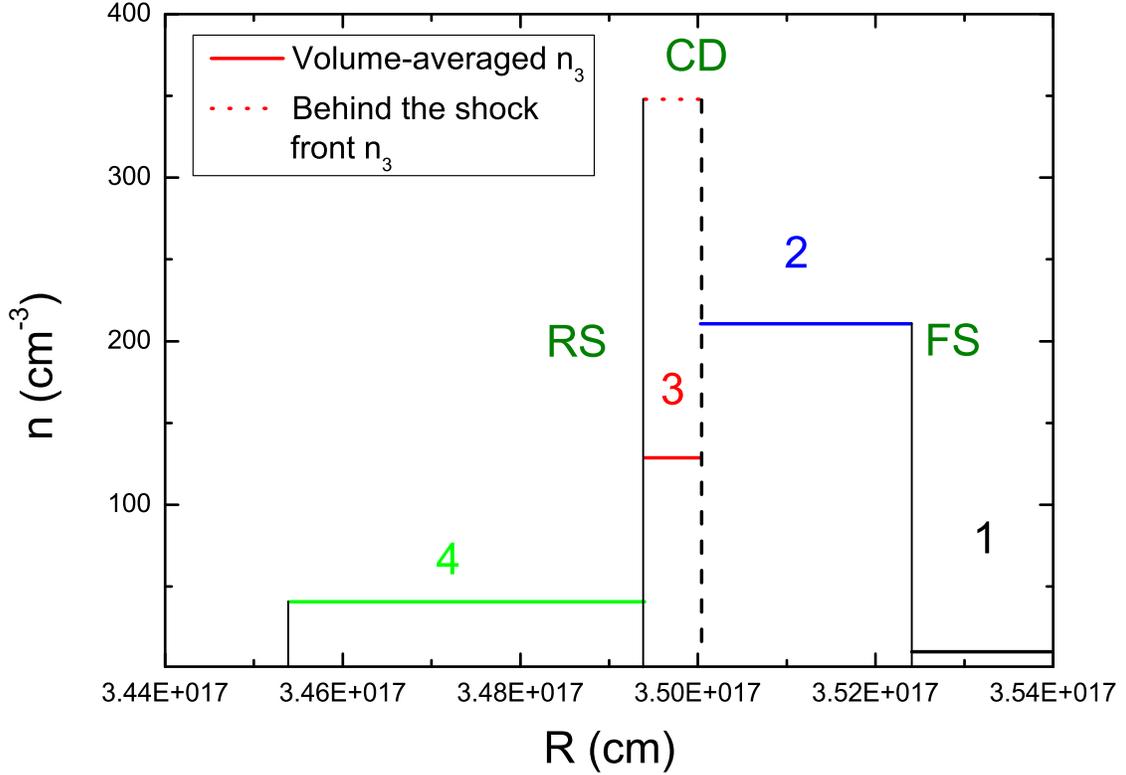}
   \caption{A sketch of the comoving density profile at the observer time $t = 40900$ s (redshift $z = 1$ is assumed) for the upper left panel in Figure 1. Region 1 refers to the ISM which the blast wave is traveling. Region 2 contains the mass swept up by forward shock (FS) after the encounter with the density change. The fluid in region 3 has been shocked by the reverse shock (RS) while region 4 contains the rest of the mass collected before the encounter. The real mean density of region 3 (red solid line) is calculated by taking the width of region 3 into account. And it is significantly lower than the value predicted by previous analytical result (red dotted line). Note that only the width of region 3 makes sense while the scales of regions 2 and 4 are unimportant in this figure.}
   \label{Fig:plot3}
   \end{center}
\end{figure}

\begin{figure}
   \begin{center}
   \includegraphics[scale=0.6]{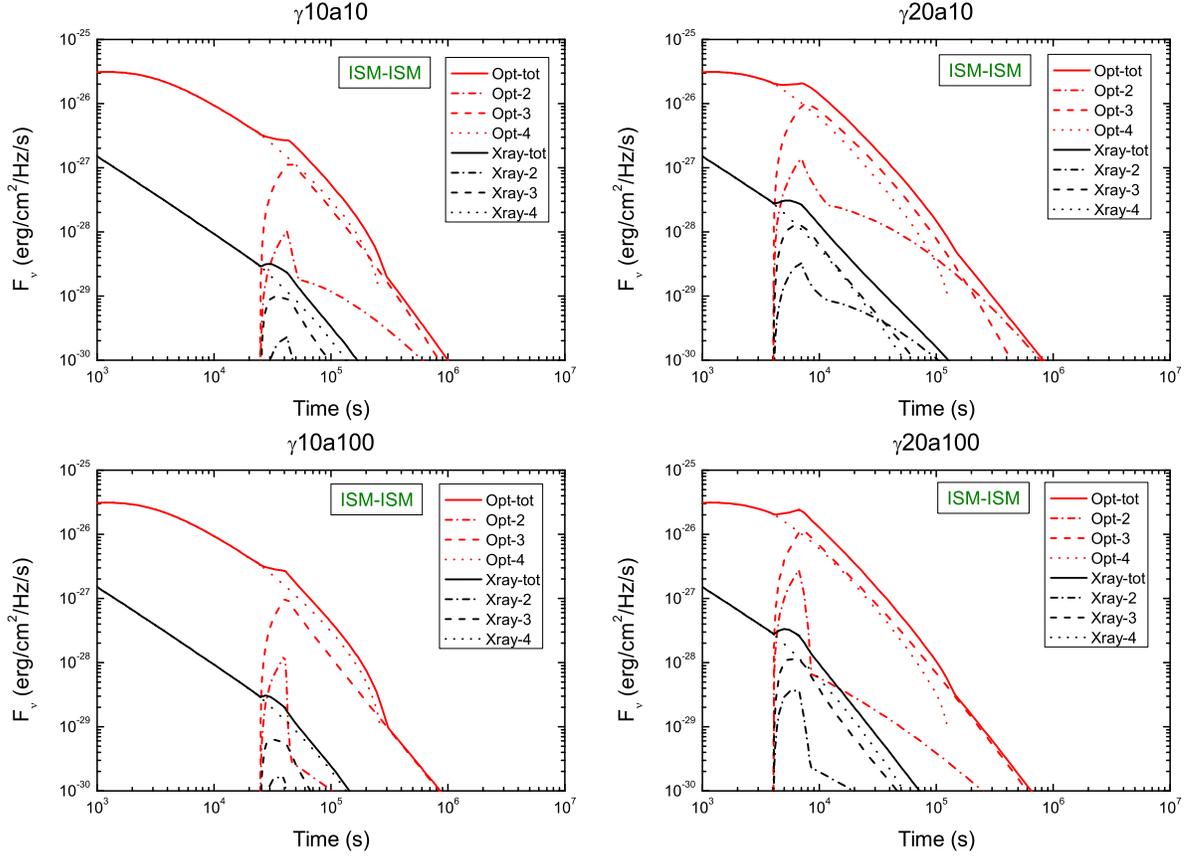}
   \caption{The corresponding afterglow light curves for the four cases of Figure 1. The red lines and black lines are the optical band ($4 \times 10^{14}$ Hz) and X-ray band ($0.3$ keV) light curves respectively. The dotted lines (Opt-4 or Xray-4) represent the flux density of region 2 before the density-jump and the flux density of region 4 after the jump. The dashed lines (Opt-3 or Xray-3) are the contribution from region 3 after the density-jump. The dash-dotted lines (Opt-2 or Xray-2) mark the flux density of region 2 after the jump, and the solid lines (Opt-tot or Xray-tot) show the total flux density of all the components. Note that the density-jump time is twice of that in Figure 1 since a redshift $z = 1$ has been assumed.}
   \label{Fig:plot2}
   \end{center}
\end{figure}

\begin{figure}
   \begin{center}
   \includegraphics[scale=0.6]{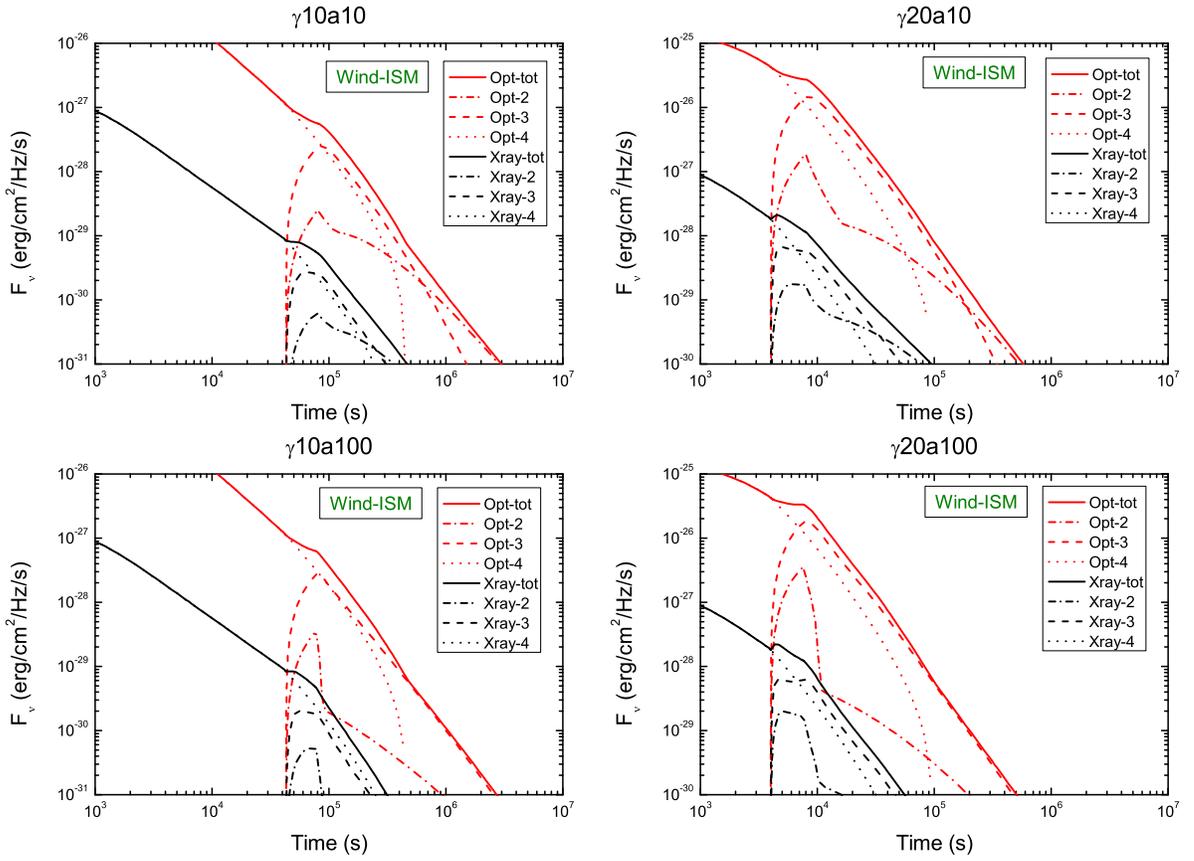}
   \caption{Same as Figure 3, but the blast wave propagates in a stellar wind environment before the encounter and enters into a homogeneous ISM after the encounter.}
   \label{Fig:plot3}
   \end{center}
\end{figure}

\begin{figure}
   \begin{center}
   \includegraphics[scale=0.3]{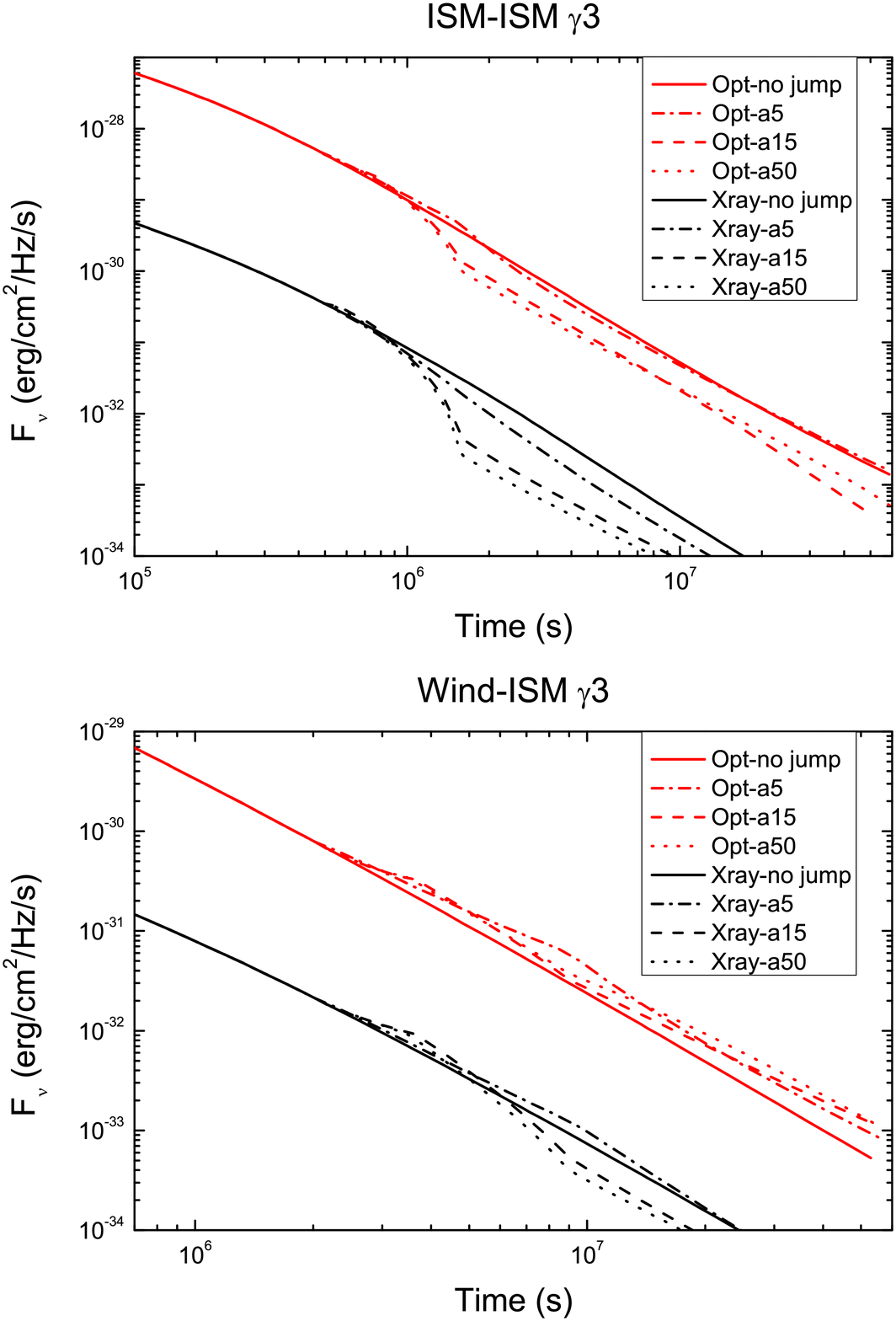}
   \caption{Figures of light curves for the blast wave traveling through an encounter of various density magnitudes ($a = 5, 15, 50$ separately) at late stage (the encountering Lorentz factor is 3). Upper panel: light curves for the case of transition from ISM to ISM. Lower panel: light curves for the case of transition from the wind environment to ISM.}
   \label{Fig:plot3}
   \end{center}
\end{figure}

\end{document}